\documentclass[11pt]{article}

\usepackage{graphicx}
\usepackage{amsmath,amssymb}

\usepackage[body={6in,8.5in}]{geometry}

\usepackage{maths_commands}
\usepackage{epstopdf, epsfig}
\usepackage[utf8]{inputenc}
\usepackage{lineno,mathrsfs,gensymb}
\usepackage{setspace,xcolor}
\usepackage{lscape,comment,multirow,appendix}
\usepackage[utf8]{inputenc}
\usepackage{color}
\usepackage{fancyhdr}
\usepackage{natbib}

\fancypagestyle{plain}{
    \fancyhf{}
    \rhead{Kasturi Shah}
    \lhead{Geophysical Fluid Dynamics Summer School 2022 Report}
    \rfoot{$\rvert$ Page \thepage}

}

\title{Scaling with the Stars: \\ The emergence of marginal stability in low $Pr$ turbulence}
\author{Kasturi Shah}
\date{January 3, 2023}

\begin{document}

\maketitle


\section{Introduction}

Shear-driven turbulence in the stratified regions of planetary oceans, atmospheres, stellar interiors, and gas giants provides an important source of vertical transport of heat, momentum, and chemical tracers. 
Stratified turbulence in astrophysical objects differs fundamentally from geophysical turbulence
because of the Prandtl number, $Pr$, which measures the ratio of momentum diffusivity to thermal diffusivity. 
Values of $Pr$ in stably stratified geophysical systems such as Earth's atmosphere and oceans are typically 0.7 and 10, respectively.
When $Pr = O(1)$, turbulent flows ($Re \gg 1$) are always close to adiabatic, i.e., thermally non-diffusive ($Pe \gg 1$). 
In stellar radiation zones of solar-type and intermediate-mass stars, 
$Pr = 10^{-9}$-$10^{-5}$ \citep{garaud2021}.
The Peclet number, which is the ratio of the thermal diffusion timescale to the turbulent advection timescale, is also the product of the Reynolds and Prandtl number. 
With $Pr \ll 1$, $Pe \ll Re$, heat diffusion in stellar fluids is much more efficient than momentum diffusion at microscopic scales and the time scale for non-adiabatic effects may be potentially even shorter than the advective time scale ($Pe \ll 1$).
This regime is, by contrast, not possible in geophysical turbulence where $Re \gg 1$ implies $Pe \gg 1$. 

Stratified turbulence in stars is thought to be generated by horizontal shear instabilities \citep{zahn1992}.
In a horizontal shear flow, 
due to the high stratification and low viscosity, the turbulent eddies are flat and only weakly coupled in the vertical direction. 
Their characteristic vertical scale of velocity variation, $H$, is far smaller than the characteristic horizontal scale, $L$, such that the aspect ratio $\alpha = H/L \ll 1$.
Their relative motion via horizontal rotation produces vertical shear on the vertical lengthscale, $H$,
which generates vertical motion and vertical mixing. 
%
As turbulence in stars is difficult to observe, numerical simulations of strongly stratified, $Pr \ll 1$ flows yield considerable insight into the validity of the \cite{zahn1992} mechanism.
Numerical simulations at low \citep{cope2020} and high \citep{garaud2020} Péclet number exhibit vertical velocity layering, supporting \cite{zahn1992}'s horizontal shear instability mechanism for stratified stellar turbulence. 
The flows are strongly anisotropic and exhibit scale separation, as predicted. 

Scaling relationships between the aspect ratio and modified Froude number $\Fr$, the ratio of the linear wave period to the time scale of the large-scale flow \citep{lignieres1999,skoutnev2022}, characterise the interplay between the anisotropy and the stratification.
Various scaling relationships have been proposed.
At low Péclet number, the vertical velocity is the unique forcing for the buoyancy, such that $w = \nabla^2 b$ \citep{lignieres1999}.
Two proposed scaling relationships are: $\alpha \sim \Fr$ \citep{skoutnev2022} and $\alpha \sim \Fr^{4/3}$ \citep{cope2020}. 
\cite{cope2020} explain the scalings that emerged from DNS by balancing the vertical advection of vertical velocity and the thermally-constrained buoyancy term ($w \partial_z w \sim (N^2/\kappa_T) \nabla^{-2} w$), 
while
\cite{skoutnev2022} recovers the $\alpha \sim \Fr$ scaling by balancing the vertical gradient of pressure and the thermally-constrained buoyancy term, $\partial_z p \sim (N^2/\kappa_T) \nabla^{-2} w$. 
At large Péclet number (but low Prandtl number), \cite{garaud2020} finds that the vertical lengthscale varies as $Fr^{2/3}$ in numerical simulations. 
As changes in scaling relationships predict transitions between turbulent behaviours, a self-consistent and rigorous derivation of physically reliable scalings is necessary to identify various turbulent regimes.

{\color{black} 
Existing algorithms for time integrations of slow-fast quasilinear systems provide methods to solve the derived multiscale model presented here \citep{ferraro2022,michel2019}.
The central challenge is the integration of the reduced model on two timescales and two spatial scales.
The former is typically approached by solving an eigenvalue problem for the fast-varying fields and time-stepping the slowly-varying fields on the slow timescale.
The latter is addressed by considering small spatial scales only (for simplicity) and hence suppressing the large-scale derivatives.
The key insight obtained from applying these algorithms is the evolution of the growth rate, represented by the eigenvalue, which indicates the stability of the flow.
Additionally, the algorithm explicitly calculates the amplitude of the fast-varying fluctuation fields.
Their feedback on the mean flow maintains its marginal stability. 
}

Motivated by open questions regarding the validity of scaling relationships and identification of distinct turbulent regimes,
we present a formal, multiscale analysis of governing low-$Pr$ (Boussinesq) equations at low and high $Pe_b$ in the limit of strong stratification. 
Scaling relationships between the aspect ratio and modified Froude number emerge naturally from the multiscale analysis, which is supported by prior numerical simulations revealing anisotropic, scale-separated, dynamics. 
We use our analysis to assess the validity of published scaling relationships and construct a full regime diagram.

\section{Multiscale Model Development}\label{sec:multiscale}

Consider a three-dimensional, non-rotating, incompressible, stably stratified flow expressed in a Cartesian reference frame where $z$ is aligned with gravity ${\bf g} = -g {\bf e}_z$.
Let $\pp{\boldsymbol{u}}$ denote the horizontal velocity, $w$ the vertical velocity, $p$ the pressure divided by a constant reference density, and $b$ the buoyancy perturbation with respect to a linearly stratified background.  
The fluid has, in accordance with the Boussinesq approximation, a constant kinematic viscosity $\nu$, thermal diffusivity $\kappa_T$, coefficient of thermal expansion $\beta$, and a constant stratification measured by the buoyancy frequency $N$.  
The governing equations for this configuration are, 
\begingroup
\allowdisplaybreaks
\begin{subequations}\label{eqn:spiegel_veronis}
\begin{alignat}{2}
    \pd{\pp{\boldsymbol{u}}}{t} + \left( \pp{\boldsymbol{u}} \cdot  \pp{\nabla} \right) \pp{\boldsymbol{u}} + w \pd{\pp{\boldsymbol{u}}}{z} &= - \pp{\nabla} p + \nu \left( \pp{\nabla}^2 \pp{\boldsymbol{u}} + \pd{^2 \pp{\boldsymbol{u}}}{z^2} \right) + f(z) \hat{\boldsymbol{e}}_x,  \\
    \pd{w}{t} + \left( \pp{\boldsymbol{u}} \cdot \pp{\nabla} \right) w + w \pd{w}{z}  &= - \pd{p}{z} + b + \nu \left( \pp{\nabla}^2 w + \pd{^2 w}{z^2} \right), \\
    \pp{\nabla} \cdot \pp{\boldsymbol{u}} + \pd{w}{z} &= 0, \\
    \pd{b}{t} + \left( \pp{\boldsymbol{u}} \cdot \pp{\nabla} \right) b + w \pd{b}{z} + N^2 w &= \kappa_T \left( \pp{\nabla}^2 b + \pd{^2 b}{z^2} \right),
\end{alignat}
\end{subequations}
\endgroup
where the horizontal gradient operator is denoted by $\pp{\nabla}$ and $\perp$ represents the horizontal coordinates $x$ and $y$.
A body force $f e_x$ (where $f$ is a function of $z$ only) is applied to drive a mean horizontally sheared flow.

\subsection{Low Péclet number equations}

Motivated by evidence of strongly anisotropic flows in numerical simulations of thermally diffusive stellar fluids \citep{cope2020,garaud2020}, we non-dimensionalise the system in \eqref{eqn:spiegel_veronis} anisotropically by defining dimensionless (hatted) variables
\begin{equation}\label{eqn:dimless_scalings}
    (x,y) = L (\hat{x},\hat{y}), \quad z = \alpha L \hat{z}, \quad \boldsymbol{u}_\perp = U \hat{\boldsymbol{u}}_\perp, \quad t = \frac{L}{U} \hat{t}, \quad w= \alpha U \hat{w}, \quad p = U^2 \hat{p}, \quad b = \alpha^3 \frac{N^2 UL^2}{\kappa_T} \hat{b}, 
\end{equation}
where 
$U$ is the characteristic horizontal velocity scale. 
Note that the body force is of order $U^2/L$.
At low Péclet number, the vertical velocity is the unique forcing for the buoyancy, and we expect $N^2 w = \kappa_T \nabla^2 b$ \citep{lignieres1999}.
Accordingly, we chose the dimensionless scaling of $b$ in \eqref{eqn:dimless_scalings} to obtain this balance in the limit $Pe \rightarrow 0$.
On substituting the newly defined variables in \eqref{eqn:dimless_scalings} and omitting the hats, \eqref{eqn:spiegel_veronis} becomes
\begin{subequations}\label{eqn:dimensionless_eqn}
\begin{alignat}{2}
    &\pd{\pp{\boldsymbol{u}}}{t} + \left( \pp{\boldsymbol{u}} \cdot \nabla_\perp \right) \pp{\boldsymbol{u}} +  w \pd{\pp{\boldsymbol{u}}}{z} = - \pp{\nabla} p + \frac{1}{Re \alpha^2} \left( \alpha^2 \pp{\nabla}^2 \pp{\boldsymbol{u}} + \pd{^2 \pp{\boldsymbol{u}}}{z^2} \right) + f, \\
    &\pd{w}{t} + \left( \pp{\boldsymbol{u}} \cdot \nabla_\perp \right) w + w \pd{w}{z} = -\frac{1}{\alpha^2} \pd{p}{z} + \frac{\alpha^2}{Fr_M^4} b + \frac{1}{Re \alpha^2} \left( \alpha^2 \pp{\nabla}^2 w + \pd{^2 w}{z^2} \right), \\
    &\pp{\nabla} \cdot \pp{\boldsymbol{u}} +  \pd{w}{z} = 0, \\
    &\pd{b}{t} + \left( \pp{\boldsymbol{u}} \cdot \nabla_\perp \right) b + w \pd{b}{z} + \frac{1}{Pe \alpha^2} w = \frac{1}{Pe \alpha^2} \left( \alpha^2 \pp{\nabla}^2b + \pd{^2 b}{z^2} \right),
\end{alignat}
where the forcing has been non-dimensionalised by $U^2/L$.
The following dimensionless parameters arise:
\begin{equation}
    Re = \frac{UL}{\nu}, \quad \alpha = \frac{H}{L}, \quad \Fr = \left( \frac{U \kappa_T}{N^2 L^3} \right)^{1/4}, \quad Pr = \frac{\nu}{\kappa_T}, \quad Pe = Pr Re,
\end{equation}
representing the Reynolds number, the aspect ratio, the modified Froude number, the Prandtl number and the Péclet number.
Both the aspect ratio and the modified Froude number are emergent parameters.
%
Crucially, to describe low $Pe$ anisotropic flows,
\eqrefl{eqn:dimensionless_eqn}{d} requires a small \textit{buoyancy} Péclet number, $Pe_b = Pe \, \alpha^2 \ll 1$, not a low bare Péclet number. 
In this limit, \eqrefl{eqn:dimensionless_eqn}{d} reduces to
\begin{equation}\label{eqn:LPN}
    w = \alpha^2 \left( \pp{\nabla}^2 + \frac{1}{\alpha^2} \pd{^2}{z^2} \right) b.
\end{equation}
\end{subequations}
At low $Pe_b$, the vertical advection of the background buoyancy gradient is balanced by the diffusion of the individual anomaly rather than by the time tendency of the buoyancy anomaly, as at high $Pe_b$.
Alternatively, \eqref{eqn:LPN} can be derived by expanding $b$ in the Boussinesq equations in powers of $Pe$ and assuming an order unity velocity field \citep{lignieres1999}. 
At $O(Pe^0)$, the subjugation of the buoyancy to the vertical velocity emerges, i.e., $w = \nabla^2 b$, consistent with \eqref{eqn:LPN}.
\eqrefl{eqn:dimensionless_eqn}{abcf} are referred to as the low Péclet number equations (LPN). 

\subsection{Multiple scale asymptotics}

Numerical studies of strongly stratified stellar turbulence at low $Pe$ \citep{cope2020} and high $Pe$ \citep{garaud2020} exhibit anisotropy
and thus scale separation.
The dominant shear instabilities have horizontal scales commensurate with the vertical scale of variability of the large-scale flow. 
The vertical scale of velocity variation $H$ is much smaller than the horizontal scale. 
Motivated by these findings, we develop a multiscale model for low $Pe_b$ in \S 2.2 and, separately, for high $Pe_b$ flows in \S 2.3.
Based on the anisotropy described by the aspect ratio, we formally split the horizontal spatial scales into `slow' and `fast' scales, such that $\boldsymbol{x}_{\perp f} = \boldsymbol{x}_{\perp s}/\alpha$ and $\boldsymbol{x}_{\perp s} = \pp{\boldsymbol{x}}$, where subscript $f$ denotes fast and subscript $s$ denotes slow \citep{chini2022}.
Based on the time scale for horizontal shear in layers separated by distance $\alpha L$, we formally split the temporal scales into a `slow' and `fast' time scale, such that $t_f = t_s/\alpha$ and $t_s = t$.
Consequently, the partial derivatives transform as
\begin{equation}\label{eqn:two_scale_t_x}
    \pd{}{t} = \frac{1}{\alpha} \pd{}{t_f} + \pd{}{t_s} , \qquad \pp{\nabla} = \frac{1}{\alpha} \nabla_{\perp f} + \nabla_{\perp s}. 
\end{equation}
In accordance with the multiple scale asymptotic formalism, the buoyancy, pressure and velocity fields are functions of both $\boldsymbol{x}_{\perp f}$ and $\boldsymbol{x}_{\perp s}$ and of both $t_f$ and $t_s$.
For a multiscale function $q(\boldsymbol{x}_{\perp f},\boldsymbol{x}_{\perp s},z,t_f,t_s;\alpha)$, we define a fast-averaging operator $\overline{( \cdot )}$,
\begin{equation}\label{eqn:fast_avg}
    \overline{q}(\boldsymbol{x}_{\perp s},z,t_s;\alpha) = \lim_{T_f,L_x,L_y \to \infty} \frac{1}{L_x L_y T_f} \int_0^{t_f} \int_{D} q (\boldsymbol{x}_{\perp f},\boldsymbol{x}_{\perp s},z,t_f,t_s;\alpha) \mathrm{d} \boldsymbol{x}_{\perp f} \mathrm{d} t_f.
\end{equation}
where $D$ is the horizontal $\boldsymbol{x}_{\perp f}$ domain, with fast spatial periods $L_x$ and $L_y$, and $T_f$ is the fast time-integration period. 
Hence, $\overline{q}$ depends on slow variables only.
Hence, $q$ can be split into a slowly-varying and a fast fluctuation component, $q-q'=\overline{q}$.
Here, primes denote fluctuation fields, where the fast-average of the fluctuation field vanishes, i.e., $\overline{q'} = 0$.

\subsection{Multiple scale quasilinear model for low Péclet flows}

We begin with the development of the multiscale model for low $Pe_b$ flows.
We proceed to asymptotically expand the pressure, horizontal velocity, vertical velocity, and buoyancy.
The expansion proceeds in fractional powers of $\alpha$ where the exponent, $\gamma$, in the expansion is determined separately in the high $Pe_b$ and the low $Pe_b$ cases.
We posit the following asymptotic expansions,
\begin{subequations}\label{eqn:asym_expansions}
\begin{alignat}{2}
    [p,\pp{\boldsymbol{u}}] &\sim [p_0, \boldsymbol{u}_{\perp 0}] + \alpha^\gamma [p_1, \boldsymbol{u}_{\perp 1}] + \alpha^{2 \gamma} [p_2, \boldsymbol{u}_{\perp 2}] + \dots, \\
    [b,w] &\sim \frac{1}{\alpha^\gamma} [b_{-1},w_{-1}] + [b_0,w_0] + \alpha^\gamma [b_1, w_1] + \dots,
\end{alignat}
\end{subequations}
which reflect our expectation that the dominant contributions to the pressure and velocity arise on large horizontal scales; accordingly, their expansions begin at $O(1)$.
In contrast, in stratified turbulence, the vertical velocity is a small-scale field \citep{brethouwer2007,maffioli2016,cope2020}. 
For the vertical divergence of the vertical flux of horizontal momentum associated with fluctuations to feed back on the leading-order large-scale horizontal flow, the fluctuation velocities must be appropriately small, given the 3D incompressibility of the isotropic fluctuating flow. 
Specifically, the vertical divergence of the fluctuation flux is $\partial_z ( \overline{w'u'}) = O[(U'/U) (W'/U)(1/\alpha)]$ relative to the inertial terms, 
where fluctuations scales are denoted as primed capital letters. 
Since $W'=U'$ only from continuity, $(W')^2 = \alpha U^2$, i.e., $W'=\alpha^{1/2} U$, so $\gamma=1/2$ and
the vertical velocity expansion starts at $w_{-1}$.
The tight coupling between vertical velocity and buoyancy in the LPN equation \eqref{eqn:LPN}, requires the asymptotic expansion of $b$ to mimic $w$.

On substituting the two-scale derivatives \eqref{eqn:two_scale_t_x} and asymptotic expansions \eqref{eqn:asym_expansions} into the LPN equations, \eqrefl{eqn:dimensionless_eqn}{abc} and \eqref{eqn:LPN}, we obtain at lowest order

\begin{subequations}\label{eqn:lowest_order_LPN}
\begin{gather}
    \pd{\boldsymbol{u}_{\perp 0}}{t_f}  + \boldsymbol{u}_{\perp 0} \cdot \nabla_{\perp f} \boldsymbol{u}_{\perp 0} = - \nabla_{\perp f} p_0, \qquad
    \pd{p_0}{z} = 0, \qquad
    \nabla_{\perp f} \cdot \boldsymbol{u}_{\perp 0} = 0. \tag{\theequation \textit{a},\textit{b},\textit{c}}
\end{gather}
\end{subequations}
Following arguments in \cite{chini2022}, we find that $\boldsymbol{u}_{0 \perp} = \boldsymbol{\overline{u}}_{0 \perp}$ only. 
Then \eqrefl{eqn:lowest_order_LPN}{a} requires that $\nabla_{\perp f} p_0 = 0$.
This combined with fast averaging \eqrefl{eqn:lowest_order_LPN}{b}, from which we obtain
$\partial_z \overline{p}_0 = 0$, 
implies that the leading-order pressure too is independent of fast horizontal and temporal scales, i.e., $p_0 = \overline{p}_0$. 
At next order, the governing equations are,
\begin{subequations}\label{eqn:next_order_LPN}
\begingroup
\allowdisplaybreaks
\begin{alignat}{2}
    \frac{\alpha^\gamma}{\alpha} \nabla_{\perp f} \cdot \boldsymbol{u}_1' + \frac{1}{\alpha^\gamma} \pd{w_{-1}}{z} &= 0, \\
    \frac{\alpha^\gamma}{\alpha} \left( \pd{\boldsymbol{u}_{\perp 1}'}{t_f} + \boldsymbol{\overline{u}}_{\perp 0} \cdot \nabla_{\perp f} \boldsymbol{u}_{\perp 1}' \right) + \frac{1}{\alpha^\gamma} w_{-1} \pd{\boldsymbol{\overline{u}}_{\perp 0}}{z} &= - \frac{\alpha^\gamma}{\alpha} \nabla_{\perp f} p_1, \\
    \frac{1}{\alpha^{\gamma+1}} \pd{w_{-1}}{t_f} + \frac{1}{\alpha^{\gamma+1}} \boldsymbol{\overline{u}}_{\perp 0} \cdot \nabla_{\perp f} w_{-1} &= - \frac{\alpha^\gamma}{\alpha^2} \pd{p_1}{z} + \frac{\alpha^2}{\Fr^4} \frac{1}{\alpha^\gamma} b_{-1}, \\
    w_{-1} &= \left( \pd{^2}{z^2} + \nabla_{\perp f}^2 \right) b_{-1}.
\end{alignat}
\endgroup
The balance of terms in \eqrefl{eqn:next_order_LPN}{ab} implies that $\alpha^\gamma/\alpha \sim 1/\alpha^\gamma$ must be true, such that the asymptotic parameter in \eqref{eqn:asym_expansions} is $\alpha^{1/2}$.
Hence, \eqrefl{eqn:next_order_LPN}{ab} provide a mathematical basis for our expectation that $\gamma=1/2$, which arose from physical arguments about the order of magnitude of the vertical divergence of the vertical flux relative to the inertial terms.
\eqrefl{eqn:next_order_LPN}{cd} then implies a balance $\alpha^{-3/2} \sim \alpha^{3/2}/\Fr^4$, yielding the crucial scaling relationship $\alpha \sim \Fr^{4/3}$.
\end{subequations}
Fast averaging \eqref{eqn:next_order_LPN} then gives
\begin{subequations}\label{eqn:fast_avg_nextorder_LPN}
\begin{gather}
    \pd{\overline{w}_{-1}}{z} = 0, \qquad \overline{w}_{-1} \pd{\boldsymbol{\overline{u}}_{\perp 0}}{z} = 0, \qquad
    \pd{\overline{p}_1}{z} = \left( \pd{^{2}}{z^{2}} \right)^{-1} \overline{w}_{-1}. \tag{\theequation \textit{a},\textit{b},\textit{c}}
\end{gather}
\end{subequations}
From \eqrefl{eqn:fast_avg_nextorder_LPN}{a} we conclude that $\overline{w}_{-1} = 0$, provided $\overline{u}_{-1} = 0$ along any given $z$ plane.
As expected for strongly stratified flow, the leading order vertical velocity is larger on small than on large horizontal scales, 
i.e., $w_{-1} = w_{-1}'$.
Hence, \eqrefl{eqn:fast_avg_nextorder_LPN}{b} is trivially satisfied and \eqrefl{eqn:fast_avg_nextorder_LPN}{c} yields $\partial_z \overline{p}_1 = 0$.
Given the tight coupling between the vertical velocity and buoyancy in \eqref{eqn:LPN}, \eqrefl{eqn:fast_avg_nextorder_LPN}{a} implies that $b_{-1} = b_{-1}'$, only.
We obtain the governing equations for the fluctuations by subtracting \eqref{eqn:fast_avg_nextorder_LPN} from \eqref{eqn:next_order_LPN}.

To derive the mean flow equations, we collect terms at $O(1)$ in our asymptotically expanded equations:
\begin{subequations}\label{eqn:LPN_O1}
\begin{alignat}{2}
    \pd{\boldsymbol{u}_{\perp 2}}{t_f} + ( \boldsymbol{\overline{u}}_{\perp 0} \cdot \nabla_{\perp f} ) \boldsymbol{u}_{\perp 2} + w_0 \pd{\boldsymbol{\overline{u}}_{\perp 0}}{z} + \nabla_{\perp f} p_2 &=  \nonumber \\
    - \pd{\boldsymbol{\overline{u}}_{\perp 0}}{t_s} - ( \boldsymbol{\overline{u}}_{\perp 0} \cdot \nabla_{\perp s} ) \boldsymbol{\overline{u}}_{\perp 0} - \nabla_{\perp s} \overline{p}_0 + \frac{1}{Re_b} \pd{^2 \boldsymbol{\overline{u}}_{\perp 0}}{z^2}  - ( \boldsymbol{u}_{\perp 1} &\cdot \nabla_{\perp f}) \boldsymbol{u}_{\perp 1} - w_{-1}' \pd{\boldsymbol{u}_{\perp 1}' }{z} + f, \\
    \pd{w_0}{t_f} + ( \boldsymbol{\overline{u}}_{\perp 0} \cdot \nabla_{\perp f} ) w_0 + \pd{p_2}{z} - \left( \nabla_{\perp f}^2 + \pd{^2}{z^2} \right)^{-1} w_0 &= - ( \boldsymbol{u}_{\perp 1} \cdot \nabla_{\perp f} ) w_{-1}' - w_{-1}' \pd{w_{-1}'}{z} , \\
    \nabla_{\perp f} \cdot \boldsymbol{u}_{\perp 2} + \nabla_{\perp s} \cdot \boldsymbol{\overline{u}}_{\perp 0} + \pd{w_0}{z} &= 0,
\end{alignat}
\end{subequations}
where the \emph{buoyancy} Reynolds number is $Re_b = Re \, \alpha^2$.
We have chosen to interpret the forcing as an $O(1)$ quantity.
A necessary condition for bounded behaviour of the $O(\alpha)$ fluctuation fields is that the fast average of the right-hand side of \eqrefl{eqn:LPN_O1}{a} vanishes.
On fast averaging \eqref{eqn:LPN_O1} and making use of the continuity equation \eqrefl{eqn:next_order_LPN}{a} at $O(1/\alpha^{1/2})$, we obtain equations for the leading order mean fields, $\boldsymbol{\overline{u}}_{\perp 0}$, $\overline{w}_0$, and $\overline{b}_0$.

Gathering the results of the formal multiscale asymptotic analysis, we obtain a novel two-scale model for strongly stratified, turbulent flows at low $Pe_b$, as summarised below.

\noindent \textit{Mean flow equations}
\begin{subequations}\label{eqn:lowPe_multiscale}
\begin{alignat}{2}
    \pd{\boldsymbol{\overline{u}}_{\perp 0} }{t_s} + ( \boldsymbol{\overline{u}}_{\perp 0} \cdot \nabla_{\perp s} ) \boldsymbol{\overline{u}}_{\perp 0} + \overline{w}_0 \pd{\boldsymbol{\overline{u}}_{\perp 0} }{z} &= - \nabla_{\perp s} \overline{p}_0 - \pd{}{z} \left( \overline{w_{-1}' u_1'} \right) +  \frac{1}{Re_b} \pd{^2 \boldsymbol{\overline{u}}_{\perp 0}}{z^2} + \overline{f}_0 \\
    \pd{\overline{p}_0}{z} &= 0 \\ 
    \nabla_{\perp s} \cdot \boldsymbol{\overline{u}}_{\perp 0} + \pd{\overline{w}_0}{z} &= 0
\end{alignat}

\noindent \textit{Fluctuation equations}

\begingroup
\allowdisplaybreaks
\begin{alignat}{2}
    \pd{\boldsymbol{u}_{\perp 1}'}{t_f} + ( \boldsymbol{\overline{u}}_{\perp 0} \cdot \nabla_{\perp f} ) \boldsymbol{u}_{\perp 1}' + w_{-1}' \pd{\boldsymbol{\overline{u}}_{\perp 0}}{z} &= - \nabla_{\perp f} p_1' + \frac{\alpha}{Re_b} \left( \nabla_{\perp f}^2 + \pd{^2}{z^2} \right) \boldsymbol{u}_{\perp 1}' \\
    \pd{w_{-1}'}{t_f} + ( \boldsymbol{\overline{u}}_{\perp 0} \cdot \nabla_{\perp f} ) w_{-1}' &= - \pd{p_1'}{z} + \left( \nabla_{\perp f}^2 + \pd{^2}{z^2} \right)^{-1} w_{-1}' + \frac{\alpha}{Re_b} \left( \nabla_{\perp f}^2 + \pd{^2}{z^2} \right) w_{-1}' \\
    \nabla_{\perp f} \cdot \boldsymbol{u}_{\perp 1}' + \pd{w_{-1}'}{z} &= 0
\end{alignat}
\endgroup
\end{subequations}
Note that in \eqrefl{eqn:lowPe_multiscale}{de}, formally small higher-order Laplacian diffusion terms have been added to regularize the fluctuation dynamics in the possible presence of sharp vertical gradients or critical layers, as in \citet{chini2022}.
We note that, for the dimensionless system \eqref{eqn:dimensionless_eqn},
the vertical lengthscale $\alpha L = \Fr^{4/3} L$ is, in the limit of strong stratification, so small that mean buoyancy \emph{anomalies}, i.e., departures from the imposed linear basic state profile, do not disrupt the leading-order basic state hydrostatic balance:
$\partial_z \overline{p}_0 = 0$.
As the scaling relationship $\alpha = \Fr^{4/3}$ describes the short vertical scale between horizontal eddies \cite[c.f.][]{zahn1992}, $\partial_z \overline{p}_0 = 0$ only on these short scales. 
Higher order mean pressure terms, however, do depend on mean buoyancy via gradients in mean vertical velocity, for instance, $\partial_z \overline{p}_2 = ( (\nabla_{\perp s})^2 + \partial^2_z)^{-1} \overline{w}_2 - \partial_z \overline{w_{-1}' w_{-1}'}$. 
This higher-order buoyancy dependence offers a possible path for (weak) buoyancy effects to be incorporated into the mean dynamics of the reduced order model \eqref{eqn:lowPe_multiscale}, which may be important on \emph{larger vertical} scales.
We do not pursue this path in the present study, but instead briefly outline it here for future work.
In the horizontal momentum equation, $\overline{p}_0$ can be replaced by a composite pressure, $\overline{p}_c = \overline{p}_0 + \alpha \overline{p}_2$.
For consistency, the corresponding horizontal momentum equation accurate to $O(\alpha)$ should be derived in terms of a composite horizontal velocity, $\boldsymbol{\overline{u}}_{\perp c}$.
Finally, we emphasize that buoyancy anomalies \emph{do} affect the fluctuation dynamics.

The closed system \eqref{eqn:lowPe_multiscale} tightly couples the mean flow to the fluctuations.
The mean flow modifies the fluctuation dynamics via advection by $\boldsymbol{\overline{u}}_{\perp 0}$ and by modifications of the vertical shear. 
The fluctuation equations are linear in the fluctuations themselves.
However, the fluctuations feed back non-linearly on the mean flow via the divergence of the Reynolds stress term in the horizontal momentum equation \eqrefl{eqn:lowPe_multiscale}{a}.
Therefore, the multiscale system \eqref{eqn:lowPe_multiscale} has a (generalised) quasilinear form.
A central result of this study is the emergence of this quasilinearity as a consequence of the strong stratification and the associated formal asymptotic derivation: nowhere in the multiscale expansion do we invoke nor impose quasilinearity as an \emph{adhoc} closure for the mean dynamics.

\subsection{Summary of multiple scale quasilinear model for high buoyancy Péclet flows} 

Next, we develop a multiscale model for high $Pe_b$ flows, focusing only on those points of difference with the multiscale model for low $Pe_b$. 
At high $Pe_b$, we non-dimensionalise \eqref{eqn:spiegel_veronis} using the same scalings as in \eqref{eqn:dimless_scalings}, but replace the buoyancy scaling with $b = \alpha N^2 L \hat{b}$.
At large $Pe_b$, the vertical advection of the background buoyancy gradient is balanced by the time tendency of the buoyancy anomaly rather than by the diffusion of the individual anomaly, as at low $Pe_b$. 
%
We perform a multiscale analysis of the resulting dimensionless governing equations by substituting the two-scale derivatives \eqref{eqn:two_scale_t_x}.
The asymptotic expansions used are as in \eqref{eqn:asym_expansions}, except for buoyancy. 
$b$ is no longer forced by $w$ and hence $b$ is a large-scale field; accordingly 
the asymptotic expansion for $b$ begins at $O(1)$, i.e., $b=b_0 + \alpha^\gamma b_1 + \alpha^{2\gamma}b_2$.
%
Consequently, the derivation follows \citet{chini2022},
yielding the scaling relationship $\alpha=B^{-1/2} \equiv Fr$, a well-known scaling result for strongly stratified geophysical turbulence \citep{billant2001,brethouwer2007,chini2022}.
Here, the Froude number, $Fr=U/NL$, is the ratio of the buoyancy period to the time scale of the large-scale flow.
The resulting closed, generalised quasilinear two-scale system for strongly stratified, turbulent high $Pe_b$ flows is identical to the system presented in \citet{chini2022}.
We note that in the mean flow and fluctuation buoyancy equations, 
\begin{subequations}\label{eqn:highPe_multiscale}
\begin{alignat}{2}
    \pd{\overline{b}_0}{t_s} + ( \boldsymbol{\overline{u}}_0 \cdot \nabla_s ) \overline{b}_0 + \overline{w}_0 \pd{\overline{b}_0}{z} &= - \frac{1}{Pe_b} \overline{w}_0 - \pd{}{z} \left( \overline{w_{-1}' b_1'} \right) + \frac{1}{Pe_b} \pd{^2}{z^2} \overline{b}_0, \\
    \pd{b_1'}{t_f} + \left( \boldsymbol{\overline{u}}_0 \cdot \nabla_f \right) b_1' + w_{-1}' \pd{\overline{b}_0}{z} &= - \frac{1}{Pe_b} w_{-1}' + \frac{\alpha}{Pe_b} \left( \nabla_{\perp f}^2 + \pd{^2}{z^2} \right) b_1',
\end{alignat}
\end{subequations}
an inverse buoyancy Péclet number $1/Pe_b$ multiplies the vertical velocity \citep[i.e., the first term on the RHS of equations (2.30) and (2.34) in][]{chini2022}.
However, this $1/Pe_b$ factor is an artefact of the choice of non-dimensionalisation. 

\section{Regimes Of Stratified Stellar Turbulence}\label{sec:turbulent_regimes}

One merit of the multiscale analysis detailed in \S \ref{sec:multiscale} is that scaling relationships for the aspect ratio naturally emerge.
These relationships for low $Pe_b$ and high $Pe_b$ flows are,
\begin{subequations}\label{eqn:scalings_lowPr}
\begin{gather}
    \alpha \sim \Fr^{4/3}, \qquad \alpha \sim Fr, \tag{\theequation \textit{a},\textit{b}}
\end{gather}
\end{subequations}
respectively. 
We now assess the validity of published scaling relationships.
Numerical simulations in \citet{garaud2020} imply scalings for: the vertical eddy scale $\hat{l}_z \sim Fr^{2/3}$, the root-mean square (rms) vertical velocity $\hat{w}_\text{rms} \sim Fr^{2/3}$, and the rms temperature fluctuations $\hat{T}_\text{rms} \sim Fr^{4/3}$, where we have translated their notation using $B=Fr^{-2}$.
The hatted variables are dimensionless. 
Given the $\alpha \sim Fr$ scaling in \eqrefl{eqn:scalings_lowPr}{b}, 
there would seem to be no theoretical basis for the scalings in \citet{garaud2020}.

In the low $Pe$ case, the $\alpha \sim \Fr^{4/3}$ scaling is verified by the numerical simulations in \citet{cope2020}.
Indeed, a key contribution of this study is that it provides a theoretical basis for the scaling underlying the numerical results in \citet{cope2020}.
To date, two distinct low $Pe$ scalings have been proposed in the literature, $\alpha \sim \Fr^{4/3}$ \cite[this study validated by][]{cope2020} and $\alpha \sim \Fr$ \cite[][or from hydrostatic balance in the anisotropically scaled equation \eqrefl{eqn:dimensionless_eqn}{b}]{skoutnev2022}.
A key difference between the $\alpha \sim \Fr^{4/3}$ and $\alpha \sim \Fr$ relationships is the vertical scales they describe;
for instance, our low $Pe_b$ multiscale model \eqref{eqn:lowPe_multiscale} with its intrinsic $\alpha \sim \Fr^{4/3}$ scaling describes the short vertical scales between the decoupled horizontal eddies that generate vertical shear \citep{zahn1992}. 
These differences raise a natural question: which of these regimes (if either) characterises turbulence in stars?

\begin{figure}
    \centering
    \includegraphics[width=\textwidth]{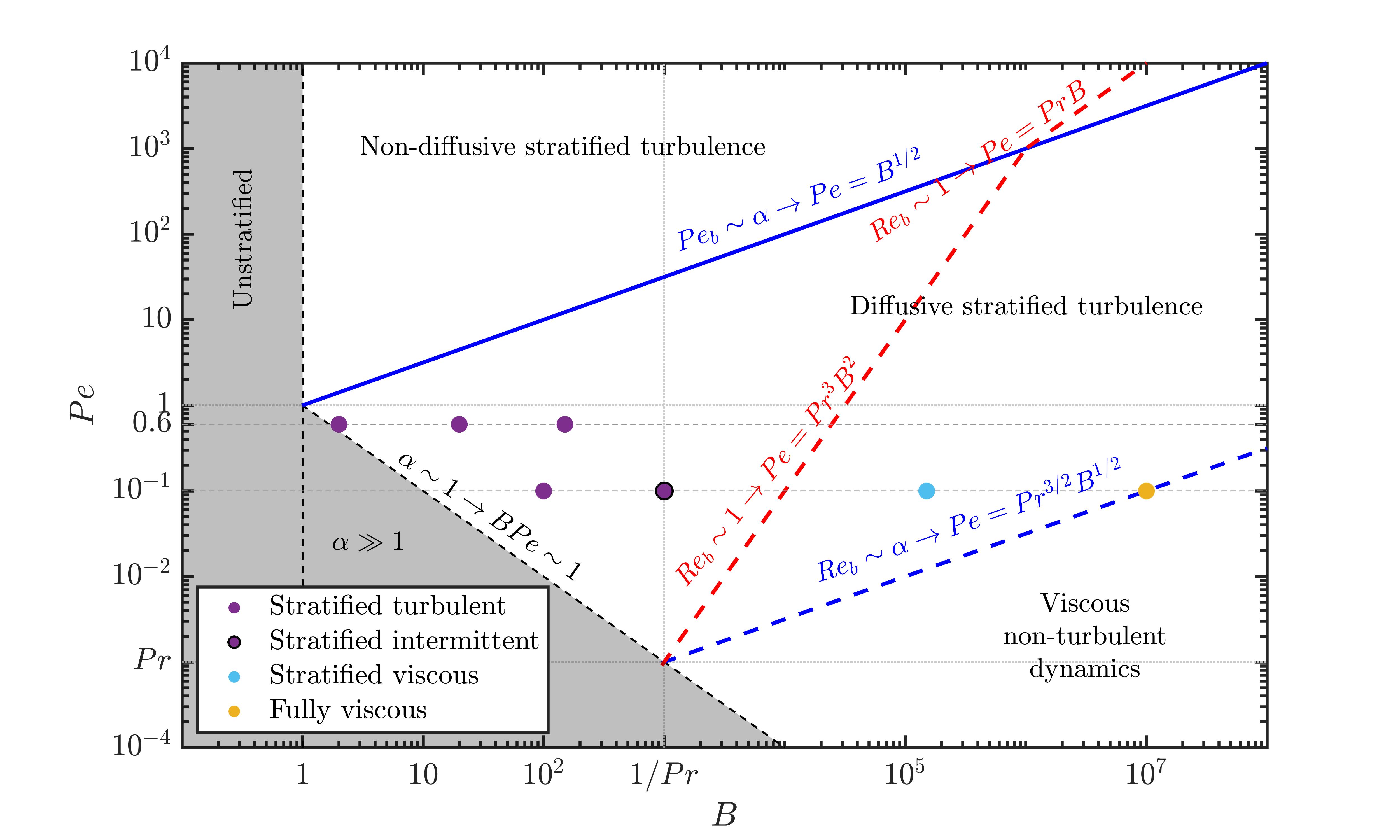}
    \caption{Regime diagram showcasing stellar turbulent behaviours for $Pr=10^{-3}$.
    The blue solid and dashed lines mark the adiabatic and viscous transitions in \eqref{eqn:lowPe_validity}.
    The low $Pe$ multiscale equations \eqref{eqn:lowPe_multiscale} are valid between these blue bounds.
    The parameter space above the blue solid line corresponds to adiabatic stratified turbulence where the high $Pe$ multiscale model applies. 
    The red dashed line marks the viscous mean flow transition in \eqref{eqn:viscous_mean}.
    The coloured circles represent numerical simulations whose behaviour is classified following \citet{cope2020}.}
    \label{fig:regime_diagram}
\end{figure}

To address this question, in Figure~\ref{fig:regime_diagram} we construct a regime diagram for stellar turbulence. 
Our multiscale models for low and high $Pe_b$ flows 
describe stratified, anisotropic flows.
The unstratified regime where our multiscale models do not apply is identified by regions where $Fr > 1$, i.e., when $B < 1$.
For $Pe_b < 1$ (where the LPN approximation is valid), the demarcating line $B Pe \sim 1$ (along which $\alpha \sim 1$) represents isotropic flows.
We first identify the viscous and adiabatic bounds between which the low $Pe_b$ multiscale model \eqref{eqn:lowPe_multiscale} is valid.
To identify the viscous transition of the fluctuation fields, we consider the multiscale fluctuation equation in the low $Pe$ case \eqrefl{eqn:lowPe_multiscale}{de}, in which, relative to the mean dynamics, viscous diffusion of fluctuation momentum is weak by the factor $\alpha$. 
Therefore, the viscous transition for the fluctuation fields occurs at $Re_b = \alpha$ and the viscous fluctuation regime arise for $\alpha/Re_b \ll 1$.
For the adiabatic transition, the relevant parameter is $Pe_b$ rather than $Pe$; 
accordingly, we consider the multiscale buoyancy fluctuation equation for low $Pe_b$ but high $Pe$ in \eqrefl{eqn:dimensionless_eqn}{d}.
The balance $\partial_t b_{-1}' + \dots = ( \alpha/Pe_b ) \nabla^2 b_{-1}$ indicates that this transition occurs at $Pe_b = \alpha$.
Adiabatic dynamics occur when $Pe_b/\alpha \gg 1$.
The range of validity of the LPN multiscale model is therefore $1/Re \ll \alpha \ll 1/Pe$.
For ease of comparison with previously published regime diagrams \citep{cope2020,garaud2020}, we express the resulting inequalities in terms of $Pe$ and $B$ using $\Fr = (B Pe)^{-1/4}$. 
On substituting the scaling relationship \eqrefl{eqn:scalings_lowPr}{a}, the region of LPN validity where the low $Pe$ multiscale model \eqref{eqn:lowPe_multiscale} applies is
\begin{equation}\label{eqn:lowPe_validity}
    Pr^{3/2} B^{1/2} \ll Pe \ll B^{1/2},
\end{equation}
demarcating where diffusive stratified turbulence occurs.
Notably, the adiabatic bound $Pe \ll B^{1/2}$ can be equivalently derived using the high $Pe$ scaling given in \eqrefl{eqn:scalings_lowPr}{b}.

Next, we identify the boundary along which the mean flow becomes viscous.
This occurs when $Re_b = O(1)$ which, on substitution of \eqrefl{eqn:scalings_lowPr}{ab}, yields 
\begin{equation}\label{eqn:viscous_mean}
    Pe \sim Pr^3 B^2, \qquad Pe \sim Pr B,
\end{equation}
respectively.
In Figure~\ref{fig:regime_diagram}, the region above the red dashed line corresponds to non-viscous dynamics,  below the blue dashed line to fully viscous non-turbulent dynamics, and 
between the red and blue dashed lines to viscous mean flow but non-viscous fluctuations.   
To corroborate the theoretical predictions, 
two sets of simulations at $Pr=10^{-3}$ are overlaid in coloured circles in Figure~\ref{fig:regime_diagram}: $ Re = 100$ and $Pe = 0.1$, which solve \eqrefl{eqn:dimensionless_eqn}{abcd}, and $Re = 600$ and $Pe=0.6$, which solve the LPN equations \eqrefl{eqn:dimensionless_eqn}{abcf}.
These simulations are categorised into stratified turbulent, stratified intermittent, stratified viscous, and fully viscous behaviours, classified consistently with \citet{cope2020}.
The stratified turbulent and intermittent simulations (purple/purple with black outline) behave independently of viscosity and lie in the non-viscous region of the diffusive stratified turbulent regime.
The stratified viscous simulation (cyan) is in the region of the diffusive stratified regime with a viscous mean flow and non-viscous fluctuations, while the fully viscous simulation (yellow) lies in the viscous non-turbulent regime.  
Hence, there is compelling agreement between the theoretical predictions and the independently classified numerical results

Returning to the question of which scaling relationship, $\alpha \sim \Fr^{4/3}$ or $\alpha \sim \Fr$, may be expected to be realised in stars, we consider where the latter can occur in Figure~\ref{fig:regime_diagram}.
As the lines indicating the viscous mean flow and fluctuation transitions, $Re_b \sim 1$ and $Re_b \sim \alpha$ respectively, intersect exactly at the isotropic transition $\alpha \sim 1$, there is evidently no region in the regime diagram where the $\alpha \sim \Fr$ scaling in \citet{skoutnev2022} applies.
We note that his scaling is only dynamically consistent with the strongly stratified $\alpha \sim Fr$ relationship when there are no small scales and when $\Fr = Fr$ interchangeably.

{\color{black}

\section{Marginal Stability Of Vertical Shear Instabilities In Low $Pe$ Flows}\label{sec:solutions}

\subsection{Time integration of the slow-fast quasilinear system}

Having established the regimes of strongly stratified stellar turbulence, we now present solutions of the slow-fast quasilinear system \eqref{eqn:lowPe_multiscale}. 
We consider three systems: 
the anisotropically scaled dimensionless governing equations in \eqref{eqn:dimensionless_eqn}, henceforth called direct numerical simulations (DNS), 
the quasilinear system integrated on one single timescale, henceforth called a single timescale quasilinear system (STQL),
and the multiple scale quasilinear system (MTQL).
Our focus is vertical shear instabilities.
To study them, we take a vertical slice through our cuboid of low $Pr$ fluid and henceforth consider the equations in $x$ and $z$ only. 
We assume a forcing of the form $f = 10 \cos(z)/Re_b \, \hat{e}_x$. 

\subsubsection{Single timescale formulation (STQL)}

The single timescale formulation evolves the entire system on a single timescale, which we choose to be the fast timescale.
To express \eqref{eqn:lowPe_multiscale} on a single time scale, we undo the chain rule, i.e. $\partial_{t_s} = \partial_{t_f}/\Fr^{4/3}$.
For convenience, we suppress the derivatives with respect to $x_s$ and zoom in to the small horizontal scales only. 
On assuming that the fast average is only horizontal,
\eqref{eqn:lowPe_multiscale} becomes 

\begin{subequations}\label{eqn:STQL}
\begin{alignat}{2}
    \frac{1}{\Fr^{4/3}} \pd{\overline{u}_0}{t_f} - \frac{1}{Re_b} \pd{^2 \overline{u}_0}{z^2} &= - \pd{}{z} \left( \int w_1' u_1' \mathrm{d}x \right) + \overline{f}_0, \\
    \pd{u_1'}{t_f} + \pd{p_1'}{x_f} - \frac{\Fr^{4/3}}{Re_b} \left( \pd{^2}{x_f^2} + \pd{^2}{z^2} \right) u_1' &= - \overline{u}_0 \pd{u_1'}{x_f} - w_1' \pd{\overline{u}_0}{z}, \\
    \pd{w_1'}{t_f} + \pd{p_1'}{z} - b_1'  - \frac{\Fr^{4/3}}{Re_b} \left( \pd{^2}{x_f^2} + \pd{^2}{z^2} \right) w_1' &= - \overline{u}_0 \pd{w_1'}{x_f}, \\
    \left( \pd{^2}{x_f^2} + \pd{^2}{z^2} \right) b_1' - w_1' &= 0, \\
    \pd{u_1'}{x_f} + \pd{w_1'}{z} &= 0,
\end{alignat}
\end{subequations}
where the fast and slow fields co-evolve on $t_f$.

\subsubsection{Multiple timescale stability analysis (MTQL)}\label{sec:MQTQL_algorithm}

We consider a small 2D domain with proportionate horizontal and vertical lengths (dimensionally, these lengths are of order $H$). 
We introduce a fluctuation streamfunction formulation,
\begin{equation}\label{eqn:psi}
    u_1' = \pd{\psi'}{z}, \qquad w_1' = - \pd{\psi'}{x_f}. 
\end{equation}
On substituting \eqref{eqn:psi} into \eqref{eqn:lowPe_multiscale}, 
the mean field equations become, 
\begin{subequations}\label{eqn:fluct_psi}
\begin{alignat}{2}
    \pd{\overline{u}_0}{t_s} + \left( \overline{u}_0 \cdot \nabla_s \right) \overline{u}_0 + \overline{w}_2 \pd{\overline{u}_0}{z} &= - \nabla_s \overline{p}_0 + \pd{}{z} \left( \overline{ \pd{\psi'}{z} \pd{\psi'}{x_f} } \right) + \frac{1}{Re_b} \pd{^2 \overline{u}_0}{z^2} + \overline{f}_0, \\
    \overline{b}_2 &= \left( \nabla^2_s + \pd{^2}{z^2} \right)^{-1} \overline{w}_2, \\
    \nabla_s \cdot \overline{u}_0 + \pd{\overline{w}_2}{z} &= 0,
\end{alignat}
and the fluctuation equations become, 
\begin{alignat}{2}
    \pd{}{t_f} \left( \pd{\psi'}{z} \right) + \left( \overline{u}_0 \cdot \pd{}{x_f} \right) \left( \pd{\psi'}{z} \right) - \left( \pd{\psi'}{x_f} \right) \pd{\overline{u}_0}{z} &= - \pd{p_1'}{x_f} + \frac{\alpha}{Re_b} \left( \nabla_f^2 + \pd{^2}{z^2} \right) u', \\
    \pd{}{t_f} \left( - \pd{\psi'}{x_f} \right) - \left( \overline{u}_0 \cdot \pd{}{x_f} \right) \left( \pd{\psi'}{x_f} \right) &= - \pd{p'_1}{z} - b_1' + \frac{\alpha}{Re_b} \left( \nabla_f^2 + \pd{^2}{z^2} \right) w', \\
    b_1' &= \left( \pd{^2}{x_f^2} + \pd{^2}{z^2} \right)^{-1} \left( \pd{\psi'}{x_f} \right), \\
    \pd{}{x_f} \left( \pd{\psi'}{z} \right) - \pd{}{z} \left( \pd{\psi'}{x_f} \right) &= 0.
\end{alignat}
\end{subequations}
On suppressing the $x_s$ derivatives, following \citet{chini2022}, 
the mean field equations reduce to a single equation
\begin{subequations}\label{eqn:MTQL_halfway}
\begin{equation}
    \pd{\overline{u}_0}{t_s} = \pd{}{z} \left( \overline{ \pd{\psi'}{z} \pd{\psi'}{x_f} } \right) + \frac{1}{Re_b} \pd{^2 \overline{u}_0}{z^2} + \overline{f}_0. 
\end{equation}
We eliminate the pressure in the fluctuation equations by taking $\partial_z$ of \eqrefl{eqn:fluct_psi}{d} summed with $-\partial_{x_f}$ of \eqrefl{eqn:fluct_psi}{e}, to obtain
\begin{alignat}{2}
    \left( \pd{}{t_f} + \overline{u}_0 \pd{}{x_f} \right) \left( \pd{^2}{x_f^2} + \pd{^2}{z^2} \right) \psi' &= \left( \pd{\psi'}{x_f} \right) \left( \pd{^2 \overline{u}_0}{z^2} \right) + \pd{b_1'}{x_f} + \frac{\alpha}{Re_b} \left( \pd{^2}{x_f^2} + \pd{^2}{z^2} \right)^2 \psi', \\
    b' &= -\left( \pd{^2}{x_f^2} + \pd{^2}{z^2} \right)^{-1} \left( \pd{\psi'}{x_f} \right).
\end{alignat}
\end{subequations}

We now focus on solving the slow-fast quasilinear system \eqref{eqn:MTQL_halfway} to interrogate its approach to marginal stability.
To first develop intuition before delving into the mathematical framework, consider the canonical example of self-organised criticality in which grains pour onto a flat plate from above, piling up. 
Generally over time, the grain pile is maintained at a special angle called the angle of repose.
Mini-avalanches occur intermittently to maintain this angle. 
This seemingly distinct system has a direct analogy to our stellar turbulent system, in which the \textit{amplitude of the fluctuations} (rather than mini avalanches) intermittently act to maintain the \textit{mean flow} (rather than the slope of the grains) at the \textit{stability criterion} (rather than the angle of repose).
Here, the stability criterion corresponds to the growth rate of the system being maintained at zero. 

To formulate equations which can be solved in a manner consistent with slow-fast quasilinear algorithms \citep{michel2019}, 
we first express the fluctuation streamfunction and buoyancy in separable form,
\begin{subequations}\label{eqn:separable}
\begin{alignat}{2}
    \psi'(x_f, z, t_f, t_s) &= A(t_s) \hat{\psi}(z,t_s) \exp(\sigma t_f + i k (t_s) x_f ) + \text{complex conjugate}, \\
    b'(x_f, z, t_f, t_s) &= A(t_s) \hat{b}(z,t_s) \exp(\sigma t_f + i k (t_s) x_f ) + \text{complex conjugate},
\end{alignat}
\end{subequations}
where their vertical structure is denoted by hatted variables,
the complex growth rate is
$\sigma=\sigma_r + i \sigma_i$, and $A(t_s)$ is the amplitude of magnitude $\rvert A(t_s) \rvert$. 
The Reynolds stress terms can now be cleanly written as,
\begin{equation}
    \pd{}{z} \left( \overline{ \pd{\psi'}{z} \pd{\psi'}{x_f} } \right) = \rvert A(t_s) \rvert^2 i k \left[ \pd{}{z} \left( \hat{\psi} \pd{\hat{\psi}^*}{z} - \hat{\psi}^* \pd{\hat{\psi}}{z} \right) \right] \equiv \rvert A(t_s) \rvert^2 RS. 
\end{equation}
On substituting \eqref{eqn:separable} into \eqref{eqn:MTQL_halfway}, we obtain
\begin{subequations}\label{eqn:MTQL}
\begin{alignat}{2}
    \pd{\overline{u}_0}{t_s} &= \rvert A(t_s) \rvert^2 RS + \frac{1}{Re_b} \pd{^2 \overline{u}_0}{z^2} + \overline{f}_0, \\
    \left( \sigma + ik \overline{u}_0 \right) \left( - k^2 + \pd{^2}{z^2} \right) \hat{\psi} &= ik \pd{^2 \overline{u}_0}{z^2} \hat{\psi} - ik \hat{b} + \frac{\alpha}{Re_b} \left( - k^2 + \pd{^2}{z^2} \right)^2 \hat{\psi}, \\
    \hat{b} &= \left( k^2 - \pd{^2}{z^2} \right)^{-1} ik \hat{\psi}.
\end{alignat}
\end{subequations}
We treat the fluctuation equations \eqrefl{eqn:MTQL}{bc} as a linear, autonomous eigenvalue problem.
On writing the system as a linear dynamical operator $\mathcal{L} X=0$, we obtain
\begin{subequations}
\begin{equation}\label{eqn:LX=0}
    \mathcal{L} X =
    \begin{pmatrix}
    (\sigma + i k \overline{u}_0 ) \left( \displaystyle \pd{^2}{z^2} - k^2 \right) - i k \displaystyle \pd{^2 \overline{u}_0 }{z^2} - \displaystyle \frac{\alpha}{Re_b} \left( \pd{^2}{z^2} - k^2 \right)^2 & ik \\
    -ik & k^2 - \displaystyle \pd{^2}{z^2}
    \end{pmatrix}
    \begin{pmatrix}
        \hat{\psi} \\
        \hat{b}
    \end{pmatrix}
    = 0
\end{equation}
with periodic boundary conditions in $z$. 
We define the inner product as
\begin{equation}
    ( X_1 \rvert X_2 ) = \int_0^{l_z} X_1 (z) X_2^*(z) \mathrm{d} z \quad \forall \quad (X_1, X_2).
\end{equation}
The adjoint operator $\mathcal{L}^\dag$ satisfies $(\mathcal{L} X_1 \rvert X_2) = (X_1 \rvert \mathcal{L}^\dag X_2)$ and is calculated using integration by parts to obtain
\begin{equation}
    \mathcal{L}^\dag X^\dag =
    \begin{pmatrix}
    (\sigma^* - i k \overline{u}_0 ) \left( \displaystyle \pd{^2}{z^2} - k^2 \right) - \displaystyle 2i k \pd{\overline{u}_0 }{z} \pd{}{z} - \displaystyle \frac{\alpha}{Re_b} \left( \pd{^2}{z^2} - k^2 \right)^2 & ik \\
    -ik & k^2 - \displaystyle \pd{^2}{z^2}
    \end{pmatrix}
    \begin{pmatrix}
        \hat{\psi}^\dag \\
        \hat{b}^\dag
    \end{pmatrix}
    = 0.
\end{equation}
Note that $\mathcal{L}$ is not a self-adjoint operator as $\mathcal{L} \neq \mathcal{L}^\dag$.
\end{subequations}
To obtain an expression for the temporal evolution of the growth rate with respect to the slow time, we take the time derivative of \eqref{eqn:LX=0}, 
\begin{subequations}
\begin{equation}\label{eqn:Ldxdt}
    \mathcal{L} \pd{X}{t_s} = - \pd{\mathcal{L}}{t_s} X
\end{equation}
where the slow time derivative of $\mathcal{L}$ is
\begin{equation}\label{eqn:dLdt}
    \pd{\mathcal{L}}{t_s} = 
    \begin{pmatrix}
        \left( \displaystyle \pd{\sigma}{t_s} + i k \pd{\overline{u}_0 }{t_s} \right) \left( \displaystyle \pd{^2}{z^2} - k^2 \right) - ik \displaystyle \pd{^2}{z^2} \left( \pd{\overline{u}_0 }{t_s} \right) & 0 \\
        0 & 0
    \end{pmatrix}
    + \td{k}{t_s} M.
\end{equation}
The above matrix is singular as it has a zero determinant.
In accordance with the Fredholm alternative, for \eqref{eqn:Ldxdt} to be solvable, the RHS of \eqref{eqn:dLdt} must be orthogonal to corresponding null eigenvector $X^\dag$, i.e.,
\begin{equation}
    \left( \mathcal{L} \pd{X}{t_s} \rvert X^\dag \right) = \left( \pd{X}{t_s} \left\rvert \mathcal{L}^\dag X^\dag \right) \right. = \left( \pd{X}{t_s} \rvert 0 \right).
\end{equation}
where $X^\dag = [ \hat \psi (z), \hat b (z) ]^T$.
Therefore,
\begin{equation}
    \left( \pd{\mathcal{L}}{t_s} X \rvert X^\dag \right) = 0.
\end{equation}
\end{subequations}
Substituting for mean flow equation \eqrefl{eqn:MTQL}{a} into \eqref{eqn:dLdt}, we derive a solvability condition
\begin{subequations}\label{eqn:Cs_solvability}
\begin{equation}
    C_1 \td{\sigma}{t_s} = C_2 \rvert A(t) \rvert^2 + C_3 + C_4 \td{k}{t}
\end{equation}
which describes how $\sigma$ changes with with respect to slow time.
Here, $C_4 = \partial_k \sigma$.
As long as we insist that the fastest growing mode has a zero growth rate (i.e., that $\sigma=0$ is a local maximum over k), then $\partial_k \sigma$ vanishes for the mode of interest.
The coefficients $C_1$, $C_2$, and $C_3$ are
\begin{alignat}{2}
    C_1 &= \frac{1}{ik} \int_0^{l_z} \left[ \hat{\psi}^{\dag *} \left( \pd{^2}{z^2} - k^2 \right) \hat \psi \right] \mathrm{d} z, \\
    C_2 &= \int^{l_z}_0 RS \left[ \hat \psi \left( \pd{^2}{z^2} + k^2 \right) \hat \psi^{\dag *} + 2 \pd{\hat \psi}{z}  \pd{\hat \psi^{\dag *}}{z} \right] \mathrm{d} z, \\
    C_3 &= \int_0^{l_z} \left[ \left( \overline{f} + \frac{1}{Re_b} \pd{^2 \overline{u}_0 }{z^2} \right) \left( \hat \psi \left( \pd{^2}{z^2} + k^2 \right) \hat \psi^{\dag *} + 2 \pd{\hat \psi}{z}  \pd{\hat \psi^{\dag *}}{z} \right) \right] \mathrm{d} z. 
\end{alignat}
\end{subequations}
The total temporal evolution of the real part of the eigenvalue $\sigma_r(\overline{u}_0, \partial_z \overline{u}_0, k)$ is
\begin{equation}
    \td{\sigma_r}{t_s} = \left( \pd{\sigma_r}{t_s} \right)_k + \td{k}{t_s} \left( \pd{\sigma_r}{k} \right)_{ \overline{u}_0, \partial_z \overline{u}_0 },
\end{equation}
however, the second term on the right-hand side vanishes for the mode of interest, provided that $k(t)$ corresponds to the fastest growing unstable mode.
Hence, the total temporal evolution is $\mathrm{d}_{t_s} \sigma_r = ( \partial_{t_s} \sigma_r )_k$, and simply corresponds to the evolution of the growth rate for a given wavenumber.
In fact, we can obtain the evolution of the growth rate for a given $k$ by dividing \eqrefl{eqn:Cs_solvability}{a} by $C_1$ as follows,
\begin{equation}\label{eqn:dt_sigma}
    \left( \pd{\sigma_r}{t_s} \right)_k = \text{Re } \left( \frac{C_3}{C_1} \right) - \text{Re } \left( - \frac{C_2}{C_1} \right) \rvert A(t) \rvert^2.
\end{equation}
The final crucial piece to the method of solution of slow-fast quasilinear systems in \citet{michel2019} is that once $\sigma_r = 0$, it stays zero (i.e., $\partial_{t_s} \sigma_r = 0$), thus maintaining the system at marginal stability.
On rearrangement of \eqref{eqn:dt_sigma} such that $\partial_{t_s} \sigma_r = 0$ is satisfied, we obtain an expression for the amplitude of the fluctuations that guarantees the marginal stability of the system,
\begin{equation}\label{eqn:Asq}
    \rvert A(t) \rvert^2 = 
    \begin{cases}
        \sqrt{ \text{Re } \left( \frac{C_3}{C_1} \right) \text{Re } \left( - \frac{C_2}{C_1} \right)^{-1} } \qquad &\text{if } \sigma_r = 0 \text{ and } \text{Re } \left( \frac{C_3}{C_1} \right), \text{Re } \left( - \frac{C_2}{C_1} \right) > 0, \\
        0 \qquad &\text{otherwise.}
    \end{cases}
\end{equation}
Hence, the amplitude of the fluctuations is intermittently non-zero when $\sigma_r = 0$.

\subsection{Nonlinear evolution of energy and growth rate}

The three systems, DNS, STQL, and MTQL, are solved using the Python software package Dedalus \citep{burns2020}. 
Pseudo-spectral methods in Dedalus are used with a second-order Runge-Kutta time-stepping scheme.
The three simulations are run in a vertical domain $L_z = 2 \pi$ using a Fourier basis with 128 gridpoints.
The horizontal domain length for DNS and STQL simulations is $L_x = (2 \pi/k_c) \Fr^{4/3}$ (where we consider $k_c=0.5$).
In order to ensure that the MTQL simulation captures dynamics on the same scale as the DNS and STQL simulations and to thus facilitate a direct comparison between the three simulations, we implement a wavenumber cutoff, $k_\text{cutoff} = 2 \pi \Fr^{4/3}/k_c$.
The search over wavenumbers occurs only when $k > k_\text{cutoff}$, such that the identified fastest growing mode does not correspond to large-scale dynamics not captured by the DNS and STQL simulations.
All three simulations are forced from rest, i.e., with zero initial velocity.

We compare the total energy in the simulations, given by
\begin{equation}
    E = \frac{1}{2} \int_0^{L_x} \mathrm{d} \boldsymbol{x}_s \int_0^{L_z} \mathrm{d} z \left( u^2 + \Fr^{8/3} w^2 + b^2 \right).
\end{equation}
Note that for the STQL and MTQL simulations, the reconstructed fields are used to calculate the total energy, i.e., $(u,b,w) \to (u_0 + \Fr^{2/3} u'_1, Fr^{2/3} w_{-1}', Fr^{2/3} b_{-1}')$. 
For the dimensionless parameters $\Fr = 0.1$ and $Re_b = 1$, the total energy from the three algorithms is compared in Figure~\ref{fig:energy_comparison}.
The energy in these three simulations is compared to the energy in the mean flow only scenario (termed ``MF only" in the legend) governed by
\begin{subequations}
\begin{equation}
    \pd{u}{t} = f + \frac{1}{Re_b} \pd{^2 u}{z^2},
\end{equation}
which when solved for $f = F_0 \cos(z)/Re_b$ gives an amplitude $A(t) = F_0 Re_b (1 - \exp(-k^2t/Re_b))$ and a mean flow energy $E_{MF} = A^2/4$.
\end{subequations}

\begin{figure}
    \centering
    \includegraphics[width=\textwidth]{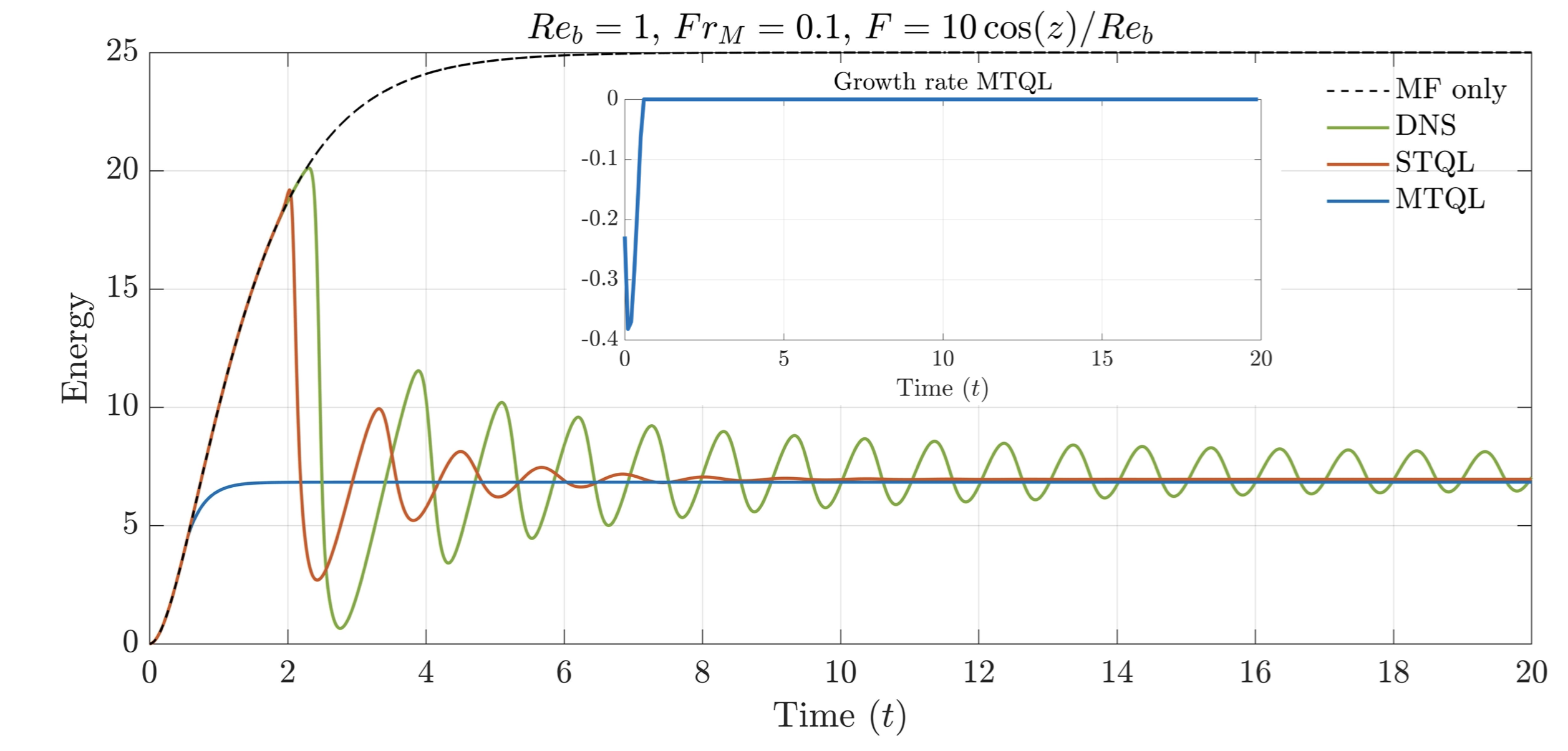}
    \caption{{\color{black}Total energy from DNS (green), STQL (red) and MTQL (blue) simulations for $Re_b = 1$ and $\Fr = 0.1$, compared against the analytical solution for a system with mean flow only (black dashed line).
    The forcing in each simulation is given by $f = 10 \cos(z)/Re_b$.
    Note that the time against which the energy is plotted here is the time of the DNS, i.e., the slow time $t_s$. 
    The inset describes the evolution of the growth rate during the MTQL simulation.}}
    \label{fig:energy_comparison}
\end{figure}

There is generally excellent agreement between the energy at which the MTQL and STQL simulations equilibrate at.
Additionally, the STQL and MTQL energy falls within the `undulations' of the energy in the DNS.
When run out for longer times (not shown), the DNS equilibrates at a similar energy to the STQL and MTQL simulations.

The three simulations fall off the mean flow curve (broken line) at different times.
As expected, the MTQL solution reaches steady-state the fastest as the fluctuation fields are instantaneously non-zero per \eqref{eqn:Asq} and adjust the mean flow instantaneously. 
The STQL solution takes longer to fall off the mean flow curve and reach steady-state as the adjustment of the fluctuation fields, while present, is not instantaneous.
The DNS simulation takes longest to reach equilibrium; this is expected as there is no formal separation of spatiotemporal scales here.

The evolution of the growth rate in the MTQL simulation is plotted in the inset of Figure~\ref{fig:energy_comparison}. 
Several key points bear discussion.
First, as the simulation is forced from rest, the real part of the growth rate over all horizontal wavenumbers is initially negative (i.e., the system is stable).
However as the flow develops with time, the growth rate approaches zero ``from below", i.e., $\sigma$ becomes increasingly less negative. 
Once $\sigma_r = 0$, the amplitude of the fluctuations in \eqref{eqn:Asq} maintains the system at marginal stability, such that from that timestep onward, $\partial_{t_s} \sigma_r = 0$.
The sudden jump of $\rvert A \rvert^2$ to a finite, non-zero value when $\sigma_r = 0$ causes a sharp change in the slow-time evolution of not just the growth rate but also the energy.

\subsection{Steady exact coherent states}

We now turn our attention from calculated quantities to the exact coherent states of the three systems.
Given the different timescales on which the DNS, STQL, and MTQL systems reach steady-state, to effect a fair comparison between the three solutions, we consider steady-state snapshots.
To identify when the simulations have reached steady-state, we plot a H{\"o}vmuller diagram for the horizontal velocity (second row in Figure~\ref{fig:snapshots}).
The simulations have clearly reached steady-state by $t=20$. 

\begin{figure}
    \centering
    \includegraphics[width=0.32\textwidth]{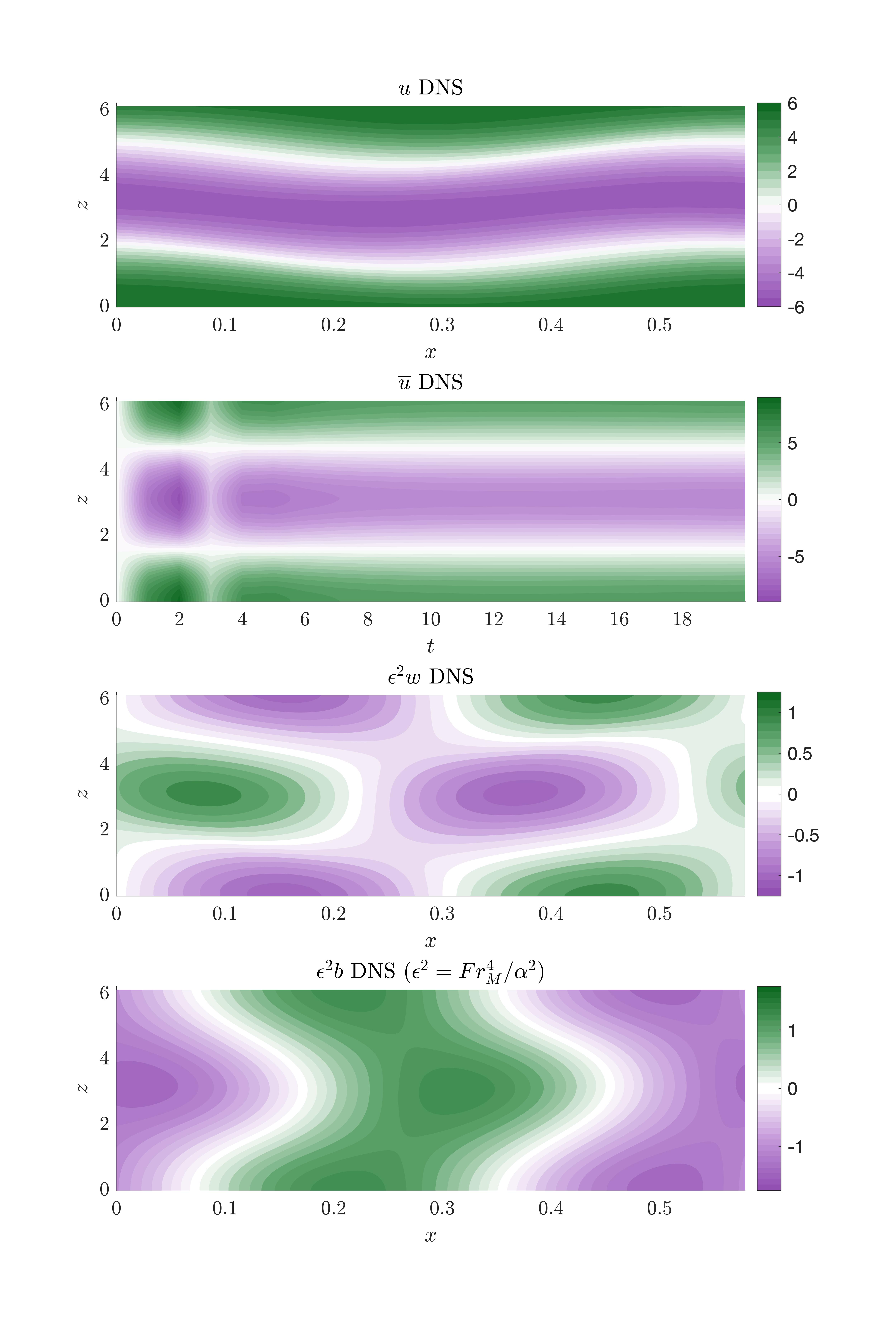}
    \includegraphics[width=0.32\textwidth]{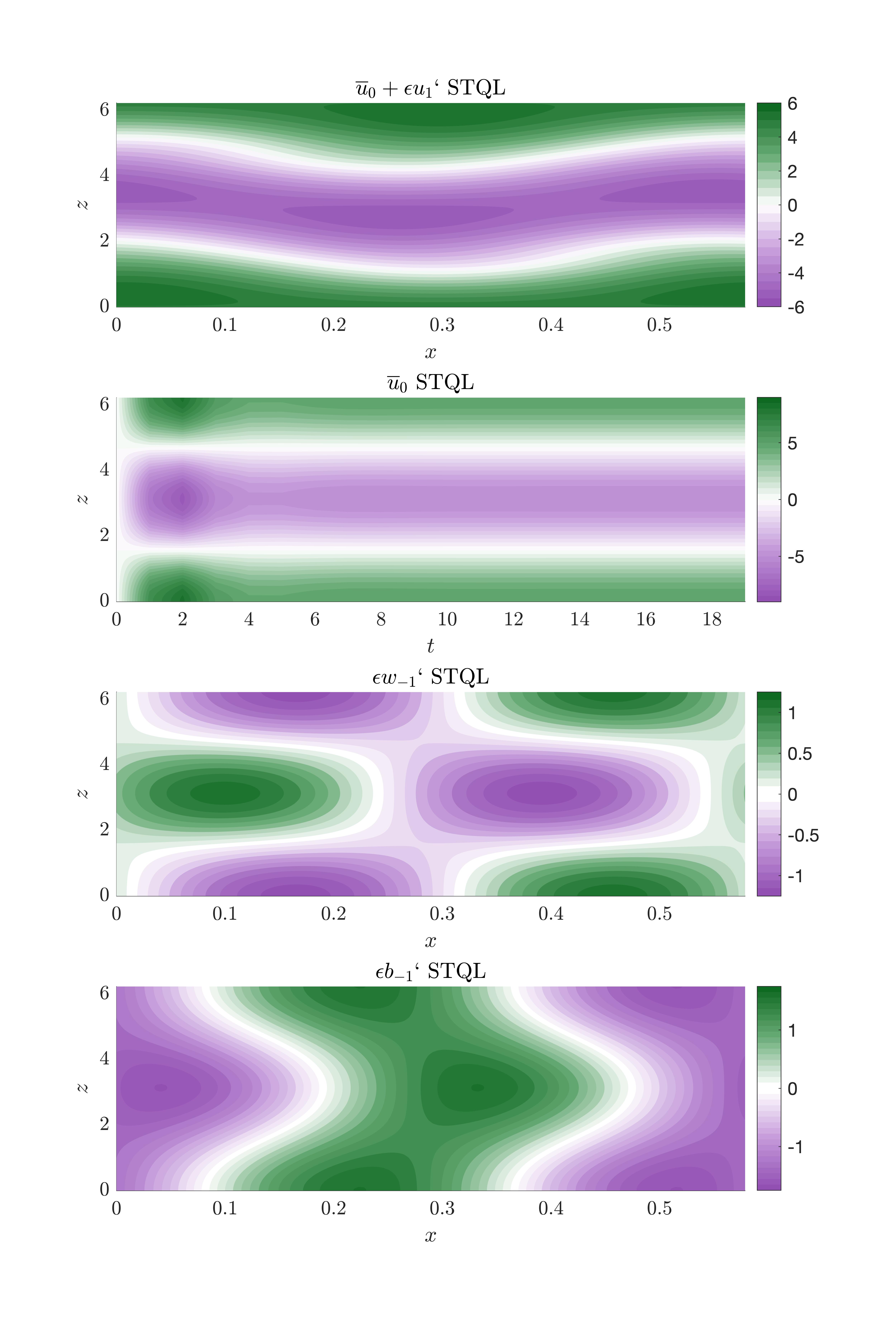}
    \includegraphics[width=0.32\textwidth]{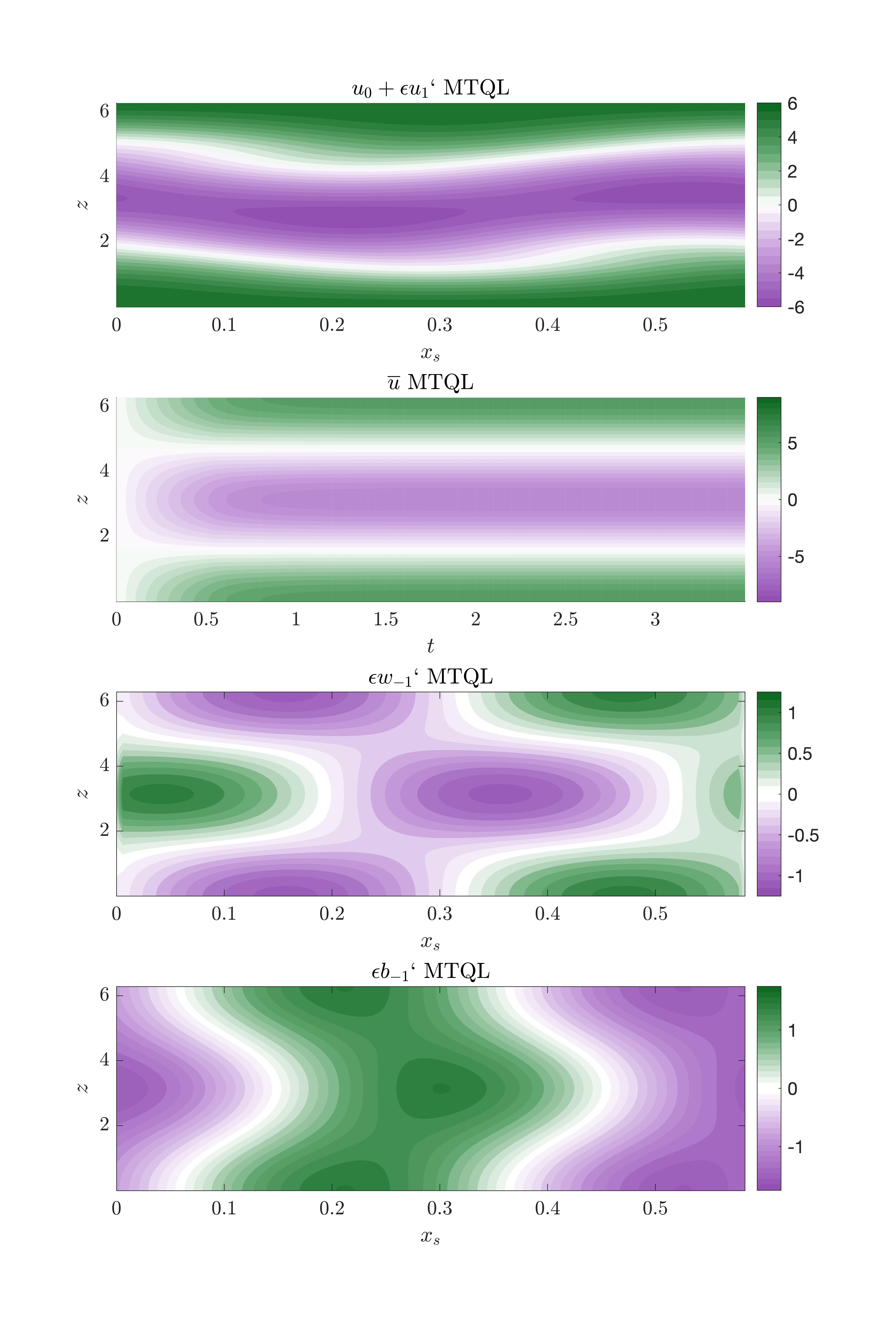}
    \caption{{\color{black}Snapshots at $t=20$ for the simulations in Figure~\ref{fig:energy_comparison}. [first row] $u$, [third row] $w$, [fourth row] $b$, plotted against $x$ ($\boldsymbol{x}_s$ for STQL and MTQL) and $z$.
    [second row] A vertically averaged H{\"o}vmuller plot is shown for $u$.}}
    \label{fig:snapshots}
\end{figure}

The horizontal, vertical and buoyancy field snapshots from all three simulations are plotted in Figure~\ref{fig:snapshots}.
For notational convenience, we introduce $\epsilon = \alpha^{1/2}$.
To compare the 2D anisotropic structure, the horizontal and vertical velocities are normalized by $U$. 
Hence, the velocities $(u,\epsilon^2 w)$ are plotted for the DNS, while $(u_0 + \epsilon u_1',\epsilon w_{-1}')$ are plotted for STQL and MTQL solutions. 
In general, there is excellent agreement between the DNS, STQL and MTQL results. 
This suggests that a quasilinear description for low Péclet flows is valid, at least for the values of $Re_b$ and $\Fr$ in Figures~\ref{fig:energy_comparison}-\ref{fig:snapshots}.


\subsection{Bursting events}

For different domain lengths, `bursting' events could arise in which the inverse ratio of coefficients in \eqref{eqn:Asq} are negative.
For notational convenience, we define
\begin{equation}
    \alpha_r = \text{Re } \left( \frac{C_3}{C_1} \right), \qquad \beta_r = \text{Re } \left( - \frac{C_2}{C_1} \right).
\end{equation}
If $\alpha_r < 0$, the positive growth rate would decay on setting $\rvert A \rvert^2=0$.
However, if $\beta_r < 0$ then the system undergoes a bursting event in which the positive growth rate increases with $\rvert A \rvert^2=0$, and even further with $\rvert A \rvert^2 > 0$.
Physically, this situation means scale separation breaks down and fast transient dynamics need to be incorporated to maintain the system at marginal stability.

Given the explosive growth associated with bursting events, how can we algorithmically solve the system when $\beta_r > 0$?
We outline several options here which apply the techniques developed for toy slow-fast quasilinear systems in \citet{ferraro2022} to the slow-fast quasilinear system for stellar turbulence, and save their algorithmic implementation for future work.
Specifically, we outline how to: (a) evolve the system to a non-bursting state, 
(b) identify when to stop this evolution, and
(c) re-initialise the MTQL system once the growth rate is negative.

From the previous timestep when $\alpha_r, \beta_r > 0$, we have the amplitude $A$, the wavenumber $k$ corresponding to the fastest growing mode, and the streamfunction $\hat{\psi}$ (from which the 2D streamfunction can be recovered by computing the outer product of $\hat{\psi}$ and $e^{ik x_f}$, where $x_f=x_s/\alpha$).
The techniques to deal with a bursting event usually involve co-evolving a system of equations until $\alpha_r, \beta_r > 0$, and then re-initialising the MTQL system. 
There are at least three possible techniques for initialising the bursting algorithm:
\begin{enumerate}
    \item Co-evolution of DNS in a 2D streamfunction formulation
    
    This approach would evolve \eqref{eqn:MTQL} on a single timescale.
    Evolution on the fast timescale would be the most judicious choice as $A$ is large, is balanced by $\partial_{t_s} \overline{u}_0$, and evolves on the fast timescale. 
    The resulting set of equations is similar to the 2D DNS system in \eqref{eqn:dimensionless_eqn}, except that it splits the spatial scale into a slow and fast component (i.e., $x=\overline{x}+x'$), and retains the eddy-eddy non-linearities.
    
    \item Co-evolution of a STQL-like system
    
    This approach involves evolving a system similar to the STQL equations in \eqref{eqn:STQL}, however, in a streamfunction formulation.
    As such, we eliminate pressure.
    The evolution considers a single wavenumber $k$, corresponding to the fastest growing mode from the previous timestep when $\alpha_r, \beta_r > 0$, to reconstruct the streamfunction. 
    
    \item A `gradient descent' strategy
    
    This approach considers the dominant balance of terms,
    \begin{equation}
        - \pd{\overline{w'u'}}{z} \sim \pd{\overline{u}_0}{t_s}
    \end{equation}
    to adopt a hybrid eigenvalue timestepping, rather than co-evolution. 
    An $O(1)$ positive number is arbitrarily assigned to the amplitude $A$, such that the amplitude is constant.
    The following equation is timestepped,
    \begin{equation}
        \pd{\overline{u}_0}{t_s} = \pd{}{z} RS
    \end{equation}
    and its eigenvalue is computed to identify the fastest growing mode.
    While more crude than the first two approaches listed above, this approach has been shown to work well for toy problems \citep{ferraro2022} and the $Pr \sim O(1)$ system of equations \citep{chini2022}.
    
    \item An appropriate rescaling strategy
    
    This approach involves finding a scaling such that the equations are free of $\epsilon$ and non-stiff.
    The resulting set of equations are then evolved on the fast timescale and without a forcing term, thus guaranteeing a reduction of the growth rate to zero.
    Ignoring the forcing term is a valid approximation when the fluctations are large, as they are in a bursting event. 
\end{enumerate}
Once the co-evolution begins, the next question to address is when to stop it.
In general, this involves solving the eigenvalue problem corresponding to the linearised dynamics, $\mathcal{L} X = 0$, as a diagnostic to monitor the (anticipated) decrease of the growth rate $\sigma_r$ towards zero during the co-evolution.
Once $\sigma_r < 0$, one would switch back to evolving the MTQL system. 

The third and final question to address is how to re-initialise the MTQL once the growth rate goes negative.
To identify the wavenumber corresponding to the fastest growing mode, the Fourier spectrum in $x$ is computed. 
(It might be necessary to vertically integrate the output from the previous timestep prior to computing the Fourier spectrum.)
Then the signs of $\alpha_r, \beta_r$ are checked and the usual conditions outlined in \S \ref{sec:MQTQL_algorithm} are applied.
As the wavenumber $k$ is discrete, when the MTQL system is re-initialised, it is possible that the wavenumber does not exactly correspond to the fastest growing mode.
There are several workarounds, such as changing the domain size with time.
Rather than implementing time-dependent coefficients, a linear transformation between coordinates can be used such that instead of a domain length $[0,L_x(t)]$, we consider a domain length of $[0,1]$. 
For instance, $x_\text{computational} = 2 \pi \, x_\text{physical}/L(t)$, and by the chain rule $\partial_{x_\text{computational}} = 2 \pi \, \partial_{x_\text{physical}}/L(t)$, where $2 \pi/L(t)$ is the wavenumber and the computational domain is $[0, 2 \pi]$.
}

\section{Conclusions}\label{sec:conclusions}

This study has established the main regimes of strongly stratified turbulence at low Prandtl number and demonstrated the approach of these turbulent flows towards marginal stability.
Stratified turbulence in stars cannot be directly measured.
Given the observational difficulties, previously published studies either simulate stellar flows at measured Froude, Prandtl and Péclet numbers \cite[e.g.][]{garaud2020,cope2020} or invoke physical arguments 
to balance terms and assess resulting scaling relationships \cite[e.g.][]{skoutnev2022}. 
The present work adopts a different approach, using numerical evidence of anisotropic flows, scale separation, and velocity layering as motivation for conducting formal, multiple scale analyses of the equations governing the dynamics of stars at low Prandtl number.
Two multiple scale models are developed, one each for low and high Péclet number flows.
A central feature of the derivation is that the generalised quasilinear form of the asymptotically-reduced equations, in which the fluctuation dynamics are shown to be linear about the mean flow and the fluctuations influence the mean flow via their induced Reynolds stress divergence, naturally emerges and is not invoked in an ad hoc fashion to close the system.
Through multiple scale asymptotics, this study provides a formal justification for the application of quasilinear approximations to descriptions of strongly stratified stellar turbulence.

The identification of distinguished limits in turbulent behaviour is a core motivation that drives the development of the multiple scale models presented here.
A second key outcome of this study is the scaling relationships for the aspect ratio that emerge via the two-scale asymptotics for high $Pe$ ($\alpha \sim Fr$) and low $Pe$ ($\alpha \sim \Fr^{4/3}$) flows. 
For low $Pe$ flows, our $\alpha \sim \Fr^{4/3}$ theoretical prediction is validated by numerical simulations in \citep{cope2020}.  
For high $Pe$ flows, our $\alpha \sim Fr$ theoretical prediction 
indicates that there is no theoretical basis for the scalings in \cite{garaud2020}.

While a star's outerscale Péclet number can be estimated from stellar observations, the emergent turbulent Péclet number can only be deduced for a given model. 
As the vast majority of stars, including our Sun, have a global scale $Pe \gg 1$ but $Pr \ll 1$, pinpointing the bound between adiabatic stratified and diffusive stratified turbulence is valuable for predicting turbulent characteristics based on stellar observations of the outerscale $Pe$.
Arguably, the primary contribution of this work is the identification of regimes of stellar turbulence.
Crucially, the momentum and buoyancy fluctuation equations in the multiscale models offer a systematic 
theoretical basis for regime identification. 
We construct a full regime diagram, identifying adiabatic stratified turbulence, diffusive stratified turbulence, and non-turbulent, viscous dynamical behaviours. 
Our theoretical identification of regimes agrees with numerical simulations, whose behaviour we classify per \citet{cope2020}.

{\color{black} 
Solutions of the multiscale slow-fast quasilinear system reveals its approach to and maintenance of marginal stability.
The system is forced from rest and hence the initial growth rate is negative but becomes increasingly less negative over time.
Once the growth rate is zero, the amplitude of the fluctuation fields acts intermittently to maintain the growth rate at zero. 
There is excellent agreement between the steady, coherent states of the DNS, STQL, and MTQL solutions.
The energy in all three systems equilibrates at similar values.
The MTQL system reaches steady-state first as the system adjusts instantaneously to the fluctuations via the divergence of the Reynolds stress restoring the system to marginal stability.
The STQL system is the second to reach steady-state as its adjustment of the mean flow by the Reynolds stress divergence is faster than the response of the fully nonlinear DNS.
}

The insight obtained from the multiscale models notwithstanding, in order to develop a truly astrophysically relevant theory for stratified stellar turbulence, physical processes such as rotation must be incorporated.
The vast majority of stars rotate.
Indeed, differential rotation typically is the main source of shear in rotating stars.
This study assumes that rotation is not needed to achieve the large horizontal scales $\boldsymbol{x}_{\perp s}$;
however, on these scales Rossby numbers are small, indicating that the large-scale dynamics are strongly affected by rotation. 
Magnetohydrodynamics too must be incorporated, given that most stars are expected to be magnetized.
Finally, a two-scale expansion in $z$ might enable the full turbulent mechanism proposed by \citet{zahn1992} to be realised, which we save for future work.

\section{Acknowledgements}

Thank you to my GFD co-advisors, Pascale Garaud, Greg Chini, and Colm-cille Caulfield, for their insight, availability, and unfailing support. 
I started the GFD school knowing very little about stars and not having researched turbulence; that I could give a coherent presentation by the summer's end is testament to their advising.
To Keaton Burns for always being open to discussing my Dedalus-related questions, no matter how small. 
To all participants of the 2022 GFD Summer School for a memorable summer and for stimulating in-person discussions that I thoroughly enjoyed after a few years of remote scientific interactions during the pandemic.
To Stefan Llewellyn Smith and Colm-cille Caulfield for their directorship and to Julie Hildebrandt and Janet Fields for their organization of the first summer school post-pandemic.
To my GFD fellows cohort (Claire Valva, Iury Simoes-Sousa, Ludovico Giorgini, Rui Yang, Ruth Moorman, Sam Lewin, Tilly Woods) for the camaraderie, the nights spent in Walsh Cottage, the swims, and the softball.
Thank you.

\bibliography{shah_gfd_report_refs}
\bibliographystyle{apalike}

\end{document}